\documentclass[prl,aps,showpacs,twocolumn]{revtex4}

\usepackage{graphicx}
\usepackage{amssymb}

\begin{document}

\title{Spin and orbital frustration in MnSc$_2$S$_4$ and FeSc$_2$S$_4$}

\author{V. Fritsch$^{1}$}
\altaffiliation[Present address: ]{Los Alamos National Laboratory,
Los Alamos, New Mexico 87545.}

\author{J. Hemberger$^{1}$}
\author{N. B\"{u}ttgen$^1$}
\author{E.--W. Scheidt$^2$}
\author{H.--A. Krug von Nidda$^1$}
\author{A. Loidl$^1$}
\author{V. Tsurkan$^{1,3}$}

\affiliation{$^1$ Experimentalphysik~V, Center for Electronic
Correlations and Magnetism,\\ Institut f\"{u}r Physik,
Universit\"{a}t Augsburg, D--86159 Augsburg, Germany}
\affiliation{$^2$ Lehrstuhl
f\"{u}r Chemische Physik und Materialwissenschaften,\\
Institut f\"{u}r Physik, Universit\"{a}t Augsburg, D--86159
Augsburg, Germany}
\affiliation{$^3$ Institute of Applied Physics,
Academy of Science of Moldova, MD 2028 Chisinau, Republic of
Moldova}

\begin{abstract}
Crystal structure, magnetic susceptibility, and specific heat were
measured in the normal cubic spinel compounds MnSc$_2$S$_4$ and
FeSc$_2$S$_4$. Down to the lowest temperatures, both compounds
remain cubic and reveal strong magnetic frustration. Specifically
the Fe compound is characterized by a Curie-Weiss temperature
$\Theta_{\rm CW}= -45$~K and does not show any indications of
order down to 50~mK. In addition, the Jahn-Teller ion Fe$^{2+}$
is orbitally frustrated. Hence, FeSc$_2$S$_4$ belongs to the rare
class of spin--orbital liquids. MnSc$_2$S$_4$ is a spin liquid
for temperatures $T>T_N\approx 2$~K.
\end{abstract}


\maketitle Frustration characterizes the inability of a system to
satisfy all pair--wise interactions and to establish unique
long--range order. Instead, a highly degenerate ground state is
formed. Frustration combined with site disorder is the key
concept to describe spin-glasses in diluted magnets \cite{Binder
86}. But frustration can also govern pure compounds due to
geometrical constraints only \cite{Ramirez 01}. Geometrical
frustration yields a variety of different ground states which
depend on the nature of the exchange interaction, the magnetic
anisotropy, and the magnitude of the spin. Spin--liquid (SL)
\cite{Canals 98} or spin--ice states \cite{Ramirez 99,Bramwell
01}, spin--clusters or --loops \cite{Garcia 00,Lee 02}, as well
as singlet formation \cite{Berg 03} were experimentally observed
or theoretically predicted.

Geometric frustration has been intensively investigated in the
spin sector. However, frustration of the orbital degrees of
freedom will be even more dominant: Orbital order is established
via the Jahn--Teller (JT) effect \cite{Jahn 37} or via a purely
electronic exchange mechanism, as proposed by Kugel and Khomskii
\cite{Kugel 82}. The exchange mechanisms in orbitally degenerate
systems are strongly frustrated and even in cubic lattices the
orbitals may remain disordered down to $T = 0$~K forming an
orbital liquid (OL) state \cite{Feiner 97,Ishihara 97,Khomskii
03}. An OL has been proposed for LaTiO$_3$ \cite{Keimer
00,Khalliulin 00} and a spin--orbital liquid (SOL) for LiNiO$_2$
\cite{Kitaoka 98}, but new experiments suggested an orbitally
ordered state in LaTiO$_3$ \cite{Cwik 03,Fritsch 03} and
questioned the SOL for LiNiO$_2$ \cite{Mostovoy 02}. Hence it
remains a challenge to search for further realizations of a SL,
OL and SOL.

Here we present experimental results on MnSc$_2$S$_4$ and
FeSc$_2$S$_4$ revealing strong spin--frustration effects.
Additionally, the Fe ion is JT--active and due to its
geometrically frustrated network, FeSc$_2$S$_4$ is expected to be
also orbitally frustrated. We thus have the interesting case to
compare two similar systems: the Mn compound, characterized by a
half--filled 3$d$ shell and zero orbital moment, being a prime
candidate for a pure SL, while the Fe compound is expected to
bear the characteristics of a SOL.

\begin{figure}
\includegraphics[width=65mm,clip,angle=0]{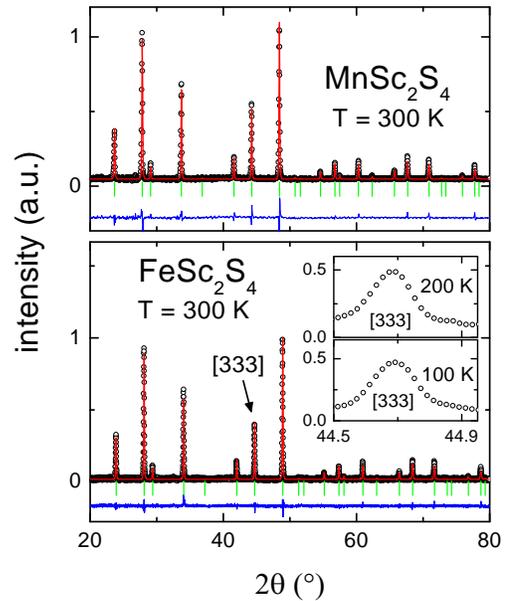}
\caption{Diffraction pattern of MnSc$_2$S$_4$ (upper frame) and
FeSc$_2$S$_4$ (lower frame). The solid lines represent fits of a
Rietveld analysis. The difference spectra shown below the data
demonstrate the absence of any impurity phases. Inset: [333]
reflection at 100 and 200~K, respectively. \label{Fig1}}
\end{figure}

It is well known that the B--sites in the normal spinels
AB$_2$X$_4$ form a pyrochlore lattice which is a paramount example
for geometrically frustrated networks. Here we demonstrate that
A--site spinels also reveal frustration, in the spin as well as
in the orbital sector. The A-site ions in the normal spinel form
a diamond lattice, i.e. two face centered cubic (fcc) lattices at
(0,0,0) and (1/4,1/4,1/4). The magnetic superexchange
interactions between the A ions are weakly antiferromagnetic and
the corresponding exchange paths involve at least five ions
\cite{Roth 64}. Within one fcc sublattice, the twelve nearest
neighbours (NN) are connected via two equivalent A-X-B-X-A
exchange paths, including nearly rectangular X-B-X bonds of
non--magnetic ions. Considering the entire lattice, the exchange
between the two A--site sublattices is transfered as follows: the
four NN are connected via six A-X-B-X-A exchange paths again
including nearly rectangular X-B-X bonds. Twelve third--next
nearest neighbours (third--NNN) are coupled via one A-X-B-X-A
exchange path, including a 180$^{\circ}$ X-B-X bond. Note that
the NNN in the entire lattice corresponds to the NN within each
sublattice. We conclude that each fcc sublattice is coupled
antiferromagnetically and, hence, is frustrated. In addition, the
two sublattices are coupled again antiferromagnetically, strongly
enforcing the frustration effects.

The synthesis of MnSc$_2$S$_4$ and FeSc$_2$S$_4$ is described in
Ref.\onlinecite{Wojtowicz 69,Tomas 79,Reil 02}. The magnetic
susceptibility for $4.2<T<300$~K was investigated by Pawlak and
Duczmal \cite{Pawlak 93} and the absence of long--range magnetic
order in both compounds has been noted. A structural
investigation \cite{Reil 02} using single crystalline
MnSc$_2$S$_4$ and polycrystalline FeSc$_2$S$_4$ confirmed the
normal spinel structure (space group $Fd\bar{3}m$) with Mn(Fe)
and Sc occupying only A-- and B--sites, respectively. Already
from earlier M\"ossbauer experiments on FeSc$_2$S$_4$ it had been
concluded that the iron ions occupy the A--sites only and that
any long--range magnetic or JT--ordering is absent \cite{Brossard
76}. The crystal field splits the $d$--electron manifold of the
divalent cations A$^{2+}$ at the tetrahedral sites into an
excited triplet ($t_{2}$) and a lower doublet ($e$).  The
Mn$^{2+}$ ion reveals a half--filled $d$--shell with a spin value
$S = 5/2$ and zero orbital moment. The Fe$^{2+}$ ion with $S = 2$
exhibits a hole in the lower doublet and, hence, is JT--active.

Polycrystalline MnSc$_2$S$_4$ and FeSc$_2$S$_4$ were prepared by
sintering stoichiometric mixtures of the pure elements in
evacuated, sealed silica ampoules at 900$^\circ$C during five
days. To reach good homogeneity the synthesis was repeated
several times with subsequent regrinding, pressing and firing.
The powdered samples were investigated by standard x-ray
techniques using Cu$K_\alpha$ radiation. Representative
room--temperature spectra are shown in figure \ref{Fig1}. From a
detailed Rietveld refinement (solid lines in Fig.\ref{Fig1}) the
lattice constant $a$ and the fractional coordinate $z$ of the
sulphur atom were determined as $a=(10.621 \pm 0.007)$~\AA ,
$z=0.2574 \pm 0.0005$ for the Mn compound and $a=(10.519 \pm
0.007)$~\AA, $z=0.2546 \pm 0.0005$ for the Fe compound,
respectively. The deviation of the sulphur parameter $z$ from the
ideal value 1/4 indicates a slight trigonal distortion of the
octahedra around the B-sites, while the tetrahedra remain
undistorted. This slight trigonal distortion yields X-B-X bonds
of 92.3$^{\circ}$ for the Fe compound and 93.6$^{\circ}$ for the
Mn compound in good agreement with Ref.\onlinecite{Reil 02}. To
search for structural phase transitions in the JT--active
compound FeSc$_2$S$_4$ we performed a diffraction analysis down
to 100~K. We found no indications of a structural distortion (see
the [333] Bragg reflection in the inset of Fig.\ref{Fig1}, which
remains resolution limited at all temperatures) and even the
positions of the sulphur atoms within the unit cell remained
constant within experimental uncertainties.

The susceptibility measurements were performed with a commercial
SQUID magnetometer for temperatures $1.7 \leq T \leq 400$~K and
in external fields up to 50~kOe. The specific--heat experiments
were conducted in noncommercial setups using a quasi--adiabatic
method for $2.5 < T< 30$~K and an AC--technique for $15 < T <
200$~K in a $^4$He--cryostat. Below 2.5~K we measured in a
${}^3$He/${}^4$He dilution refrigerator with a relaxational
method.

\begin{figure}
\includegraphics[width=70mm,clip,angle=0]{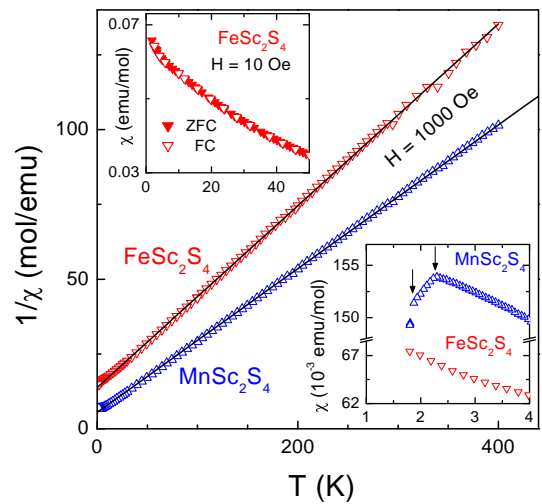}
\caption{Inverse susceptibility $1/\chi(T)$  of MnSc$_2$S$_4$
(triangles up) and FeSc$_2$S$_4$ (triangles down). The solid
lines are linear fits with a Curie--Weiss law $\chi =
C/(T-\Theta_{\rm CW})$. Upper Inset: FC and ZFC susceptibility
loop measured in FeSc$_2$S$_4$ at 10~Oe. Lower inset:
susceptibility $\chi(T)$ vs $T$ at low temperatures. \label{Fig2}}
\end{figure}

Figure \ref{Fig2} shows the inverse susceptibilities of
MnSc$_2$S$_4$ (triangles up) and FeSc$_2$S$_4$ (triangles down)
for $1.7 \leq T \leq 400$~K. We observed perfect Curie--Weiss
(CW) laws with CW temperatures $\Theta_{\rm CW} = (-45.1\pm 1)$~K
for the iron and $\Theta_{\rm CW} = (-22.9\pm 0.8)$~K for the
manganese compound, respectively. Paramagnetic moments of
$\mu_{\rm eff} = (5.12\pm 0.1)\mu_{\rm B}$ for FeSc$_2$S$_4$ and
$\mu_{\rm eff} = (5.77\pm 0.12)\mu_{\rm B}$ for MnSc$_2$S$_4$
were determined. Here we took the average of a series of
measurements of different batches, where we used as grown samples
and samples tempered in vacuum as well as in sulphur atmosphere.
Despite these different treatments, we observed marginal changes
of $\mu_{\rm eff}$ and $\Theta_{\rm CW}$ only, which are included
in the error bars given above. For Fe$^{2+}$ (3$d^6$, high spin),
the experimentally determined value of $\mu_{\rm eff}$ is higher
than the spin--only value of 4.90$\mu_{\rm B}$. This signals an
enhancement due to spin--orbit coupling, resulting in an
effective $g$-value of 2.09, typically observed in iron compounds
with Fe$^{2+}$.

A closer look at the susceptibilities $\chi(T)$ reveals slight
deviations from a CW law below 4~K in the Fe compound, but the
absence even of any spin--glass ordering is demonstrated by field
cooled (FC) and zero--field cooled (ZFC) cycles of $\chi(T)$ at
low fields (10~Oe) revealing no significant splitting (see upper
inset of Fig.~\ref{Fig2}). In MnSc$_2$S$_4$ the clear onset of
AFM order is demonstrated in the lower inset of figure
\ref{Fig2}. There, $\chi(T)$ exhibits a peak at the ordering
temperature $T_{\rm N1} = 2.1$~K and an additional downturn for
$T<T_{\rm N2} = 1.8$~K.

\begin{figure}
\includegraphics[width=65mm,clip]{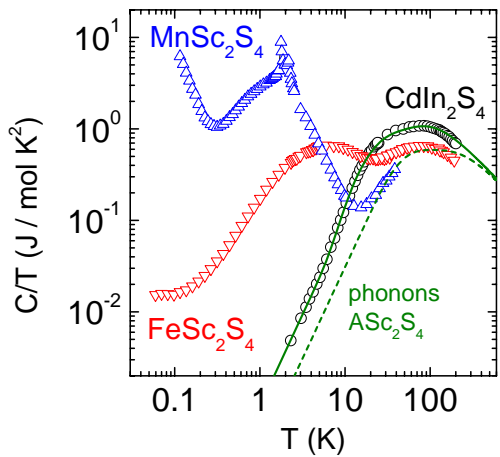}
\caption{Specific heat $C(T)/T$ for MnSc$_2$S$_4$ (triangles up),
FeSc$_2$S$_4$ (triangles down), and CdIn$_2$S$_4$ (circles). The
solid line represents the calculated specific heat of the
nonmagnetic reference compound CdIn$_2$S$_4$. The dashed line
gives the estimated phonon contribution for ASc$_2$S$_4$ (A = Mn,
Fe). \label{Fig3}}
\end{figure}

Figure ~\ref{Fig3} documents the molar heat capacity for
temperatures $0.05 < T < 200$~K. As the results cover almost four
decades in temperature, for representation purposes we plotted
$C/T$ vs $T$ on logarithmic scales. The results are compared to
nonmagnetic CdIn$_2$S$_4$ to get an estimate of the phonon
contributions. Below 200~K, in the Fe and Cd compound no
specific--heat anomalies can be detected, indicating the absence
of structural (CdIn$_2$S$_4$) or structural/magnetic
(FeSc$_2$S$_4$) phase transitions. Together with the diffraction
pattern (lower frame of Fig.\ref{Fig1}) it is clear that
FeSc$_2$S$_4$ remains cubic (no JT distortion) and paramagnetic
down to 50~mK. In all samples the lattice--derived specific heat
dominates above 20~K. However, it is impressive to see, how in
the geometrically frustrated systems the heat capacity is
enhanced towards low temperatures, having in mind that the
characteristic magnetic temperatures are 23~K and 45~K for the Mn
and Fe compound, respectively. For MnSc$_2$S$_4$, a double peak
close to 2~K in $C(T)/T$ indicates two subsequent magnetic phase
transitions which also have been observed in $\chi (T)$ (cf.
lower inset of Fig.\ref{Fig2}). The magnetic order of
MnSc$_2$S$_4$ most probably results from residual magnetic
interaction only, as any coupling to the lattice in this
spin--only system must be negligible. Below 0.4~K, nuclear
contributions dominate and the increase in $C(T)/T$ can be
accounted for by assuming a hyperfine term of $^{45}$Sc in an
internal field of 15~kOe due to the magnetic order of the
manganese moments.

\begin{figure}
\includegraphics[width=53mm,clip]{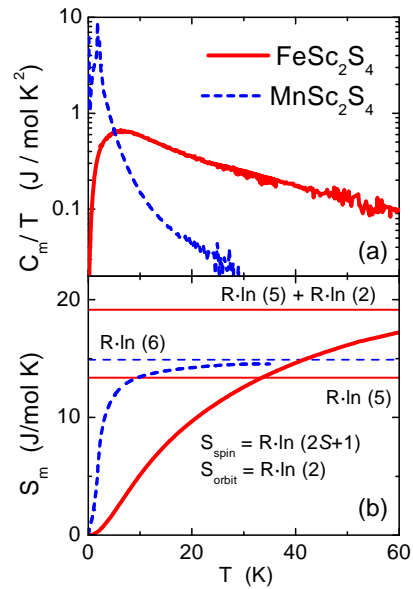}
\caption{(a) Magnetic contribution of the specific heat $C_m/T$ vs
$T$ for MnSc$_2$S$_4$ and FeSc$_2$S$_4$. (b) Magnetic entropy of
MnSc$_2$S$_4$ and FeSc$_2$S$_4$. The horizontal lines give the
magnetic entropy $S_{\rm m}$ which is theoretically expected for
the spin and orbital degrees of freedom (see text). \label{Fig4}}
\end{figure}

FeSc$_2$S$_4$ reveals a broad peak close to 6~K which often is
observed in geometrically frustrated magnets \cite{Ramirez 01}
indicating soft collective modes. On further decreasing
temperatures, $C(T)/T$ decreases reaching a constant value below
0.2~K. The lack of any hyperfine contribution signals the absence
of internal magnetic fields. The constant value of $C(T)/T$ below
0.2~K suggests a linear term in $C(T)$ which often is observed in
spin--glasses \cite{Binder 86}. The increase of the heat capacity
between $0.3 < T < 2$~K is close to $T^{2.5}$, a power--law
behavior that also has been observed in LiNiO$_2$ another
potential candidate for a SOL. As documented in figure \ref{Fig3},
FeSc$_2$S$_4$ does not order down to 50~mK, despite the fact that
Fe reveals a strong spin--spin interaction and is JT--active.
Hence, the ground state indeed has to be characterized as SOL. In
this case the magnetic (spin) frustration parameter $f = -
\Theta_{\rm CW}/T_{\rm N}$ is of the order of 1000 and one of the
largest values reported so far \cite{Ramirez 01}. In
MnSc$_2$S$_4$, the magnetic ordering temperature of 2~K and the
CW temperature of $-23$~K yields $f = 11.5$.

In order to get an estimate of the magnetic contribution to the
specific heat we calculated the lattice specific heat of
CdIn$_2$S$_4$ and scaled it to the two magnetic compounds under
investigation. The phonon contribution of CdIn$_2$S$_4$ can well
be parameterized (solid line in Fig.~\ref{Fig3}) using a Debye
term and three Einstein contributions, with a Debye temperature of
154~K and Einstein temperatures of 90~K, 244~K, and 350~K. This
assumption is corroborated by the fact that in CdIn$_2$S$_4$ the
lowest infrared(IR)--active mode is at 69 cm$^{-1}$ (99~K), while
the highest IR mode was detected at 304 cm$^{-1}$ (438 K)
\cite{Lutz 83}. In this simulation we fixed the number of
internal degrees of freedom to 21, according to seven atoms in
the unit cell. We also assumed a reasonable weight distribution
between these modes. Keeping this weight distribution fixed, we
tried to describe the phonon contribution of MnSc$_2$S$_4$ and
FeSc$_2$S$_4$ which should be practically identical due to the
similar masses of Mn and Fe. The dashed line in Fig.~\ref{Fig3}
was calculated using a Debye temperature of 190~K and Einstein
temperatures of 190~K, 420~K, and 550~K. This seems to be
reasonable taking into consideration the much lower masses of
Fe/Mn and Sc as compared to Cd and In, respectively.

Using this phonon contribution we estimated the purely magnetic
contributions $C_{\rm m}$ of the heat capacity. The results are
shown in Fig.~\ref{Fig4}a as $C_{\rm m}/T$ vs $T$ up to 60~K in a
semilogarithmic representation. We see that above 30~K the
magnetic contribution vanishes for MnSc$_2$S$_4$, while even at
60~K there are some finite contributions in $C_{\rm m}/T$ for
FeSc$_2$S$_4$. Due to the experimental and model dependent
uncertainties we confine the discussion to temperatures  $T<
60$~K, where our results are significant.

From $C_{\rm m}/T$ we calculated the magnetic entropy $S_{\rm m}$
which is plotted in figure~\ref{Fig4}b. We found that for
MnSc$_2$S$_4$ only 30\% of the expected entropy for a spin $S =
5/2$ system are recovered at $T_{\rm N}$. The full entropy is
reached at $T \approx -\Theta_{\rm CW}$. For FeSc$_2$S$_4$ the
entropy slowly increases with increasing temperature reaching
$S_{\rm m}$=Rln(5) corresponding to a $S = 2$ system at $T
\approx 30$~K. It further increases significantly towards $S_{\rm
m}$=Rln(5) + Rln(2) where the latter term characterizes the
entropy of the orbital doublet of the JT--active $e$ levels. From
the entropy $S_{\rm m}$ in Fig.~\ref{Fig4}b we attribute
MnSc$_2$S$_4$ to a SL for temperatures $2 < T < 23$~K, whereas in
the case of FeSc$_2$S$_4$ an additional contribution to $S_{\rm
m}$ comes from the frustrated orbital state. Therefore,
FeSc$_2$S$_4$ can be considered as a SOL for $T<45$~K.

In conclusion, we investigated the thiospinel systems
MnSc$_2$S$_4$ and FeSc$_2$S$_4$ and found strong frustration
effects. MnSc$_2$S$_4$ is a spin--only system with $S=5/2$,
characterized by a spin--frustration parameter $f\approx 10$
showing AFM order at low temperatures and bearing the
characteristics of a spin liquid between $T_{\rm N}\approx 2$~K
and $-\Theta_{\rm CW} \approx 23$~K. FeSc$_2$S$_4$ ($S = 2$,
JT--active) reveals frustration both in the spin and in the
orbital sector. The spin--frustration parameter $f > 900$ is one
of the largest ever observed. The orbital frustration is
evidenced by the entropy which significantly exceeds the spin
value of Rln(5). FeSc$_2$S$_4$ has to be characterized as
spin--orbital liquid below 45~K. In this case, any kind of spin
order, including canonical spin-glass behavior can be excluded on
the basis of the temperature dependence of the FC and ZFC
susceptibilities and heat capacity.

The question remains why these two similar systems behave so
differently: This problem can only be tackled taking into account
the orbital degrees of freedom coupled to the spins. MnSc$_2$S$_4$
is a spin-only system with no orbital degrees of freedom,
supported by preliminary ESR experiments yielding no indication
of any anisotropy in the resonance absorption. At some finite
temperature some residual exchange interactions (e.g. NNN
interactions or magnetic dipolar interactions) establish
long-range magnetic order. FeSc$_2$S$_4$ exhibits orbital
degeneracy and our results demonstrate indeed \cite{Feiner 97}
that orbital degeneracy does drastically increase quantum
fluctuations and thereby suppresses long-range magnetic order.

\acknowledgments This work was supported by the BMBF via VDI/EKM,
FKZ 13N6917--A and by the Deutsche Forschungsgemeinschaft through
the Sonderforschungsbereich 484 (Augsburg).

%\bibliographystyle{apsrev}
%\bibliography{Fe1-xCuxCr2S4}

\end{document}